\documentclass[preprint,showpacs,apl]{revtex4}
\usepackage{graphicx} 
\usepackage{dcolumn} 
\usepackage{bm} 

\begin{document}
\title {Quantum tunneling of  vortices in MgB$_2$ superconductor}
\author {Y. Z. Zhang} 
\affiliation{National Laboratory for Superconductivity, Institute of  Physics \& Centre for Condensed Matter Physics, Chinese Academy of Sciences, P. O. Box 603, Beijing 100080, China}
\affiliation{Universit\'{e} Libre de Bruxelles, Physique des Solides, B-1050, Brussels, Belgium}
\author {R. Deltour}
\affiliation{Universit\'{e} Libre de Bruxelles, Physique des Solides, B-1050, Brussels, Belgium}
\author {H. H. Wen, C. Q. Jin, Y. M. Ni, S. L Jia, G. C.Che, and Z. X. Zhao}
\affiliation{National Laboratory for Superconductivity, Institute of  Physics \& Centre for Condensed Matter Physics, Chinese Academy of Sciences, P. O. Box 603, Beijing 100080, China} 
\begin{abstract}
Magnetic relaxation in a MgB$_2$ superconductor was measured. The temperature dependence of the normalized relaxation rate was determined for three different magnetic fields. By extrapolating these rates to $T=0$ K, we find that these extrapolations do not approach zero, indicating quantum tunneling of vortices in the superconductor.  A quantum correction of the relaxation rate, followed by the correction of the magnetic moment, is proposed.  Using the quantum correction, we find that $U_0$ increases with decreasing temperature and approaches a maximum at $T=0$ K.
\end{abstract}
\maketitle

Recent experimental studies of vortex dynamics of MgB$_{2}$ superconductor suggest that quantum tunneling of the vortices may strongly influence the superconducting behaviors \cite {Wen}. Thus, better undersdanding the mechanism is very important for future applications.  Magnetic relaxation technique is often employed to characterize superconductors.  During the measurement, a magnetic field is first applied to a superconductor at a fixed temperature, and then the magnitude of the magnetic moment $m(t)$ is measured \textit{vs.} time. The $m(t)$ is frequently analyzed with Anderson-Kim (AK) flux-creep model \cite {Anderson}, $m(t) \approx m_{0}[1 - (T / U_{0})\ln (t / t_{0})]$, where $U_0$ is the energy barrier of the vortex motion in the absence of a driving force, $t_0$ the reference time, and $m_0$ the magnetic moment at $t=t_0$. Deviations of the AK model are found in high $T_c$ superconductors (HTSCs), leading to a more general relation \cite {Blatter,  Yesturun}
\begin{equation} m(t)\approx  \frac{m_0}{ [1+(\mu T/U_0)\ln (t/t_0)]^{\frac{1}{\mu}}}\label{1}, \end{equation}
where the characteristic factor $\mu=-1$ for the AK model, $\mu \to 0$ for the Zeldov model \cite{Zeldov,Yesturun}, and $0<\mu$ for the collective creep model \cite{Blatter, Yesturun, Feigelman}. This relation predicts that the normalized relaxation rate 
\begin{equation}R= -\frac{d\ln m}{d\ln t}\approx \frac{T}{U_0+\mu T\ln(t/t_0)}\approx \frac{T}{U_0}[1-\frac{\mu T}{U_0}\ln(t/t_0)]\label{2}, \end {equation}
[for the diamagnetic moment or for the persistent current ($m\propto j$, where $j$ is the current density)] vanishes linearly with decreasing temperature. However, experiments showed that the relaxation rates of superconductors do not extrapolate to zero \cite{Yesturun,Campbell, Civale, Hamzic, Nowak} at $T=0$ K, indicating the decay rate being probably governed by quantum tunneling of vortices \cite {Blatter,  Yesturun, Blatter1, Blatter2, Ivlev, Smith}. 

In this study, magnetic relaxation is measured at fixed temperatures in three different magnetic fields. We find that the extrapolations of the magnetic relaxation rates do not approach zero when $T\to 0$, suggesting the existence of quantum effects in the vortex system. Based on the theory of  quantum tunneling vortices and the magnetic relaxation data, a quantum correction relation is proposed, followed by corresponding barrier evaluations. 

A small MgB$_2$ thin chip (thickness about 0.2 mm, and weight about 2.2 mg) was studied \cite{Li}.  The sample shows a sharp superconducting transition temperature around 39 K with a corresponding nearly single phase of MgB$_{2}$ as determined by X-ray diffraction. Using a magnetic property measurement system (SQUID, Quantum Design MPMS 5.5 T), the magnetic relaxation of the sample was measured in magnetic fields with a scanning length 3.0 cm. The measurements are processed at several fixed temperatures and magnetic fields at 1.0, 2.0, and 3.0 T.  The sample was first zero-field cooled to a measurement temperature from a temperature ($T=50.0$ K) well above the superconducting transition temperature $T_c$, and then a magnetic field was applied for the measurement.

\begin{figure} [t]
\includegraphics
[width=.6\columnwidth]
{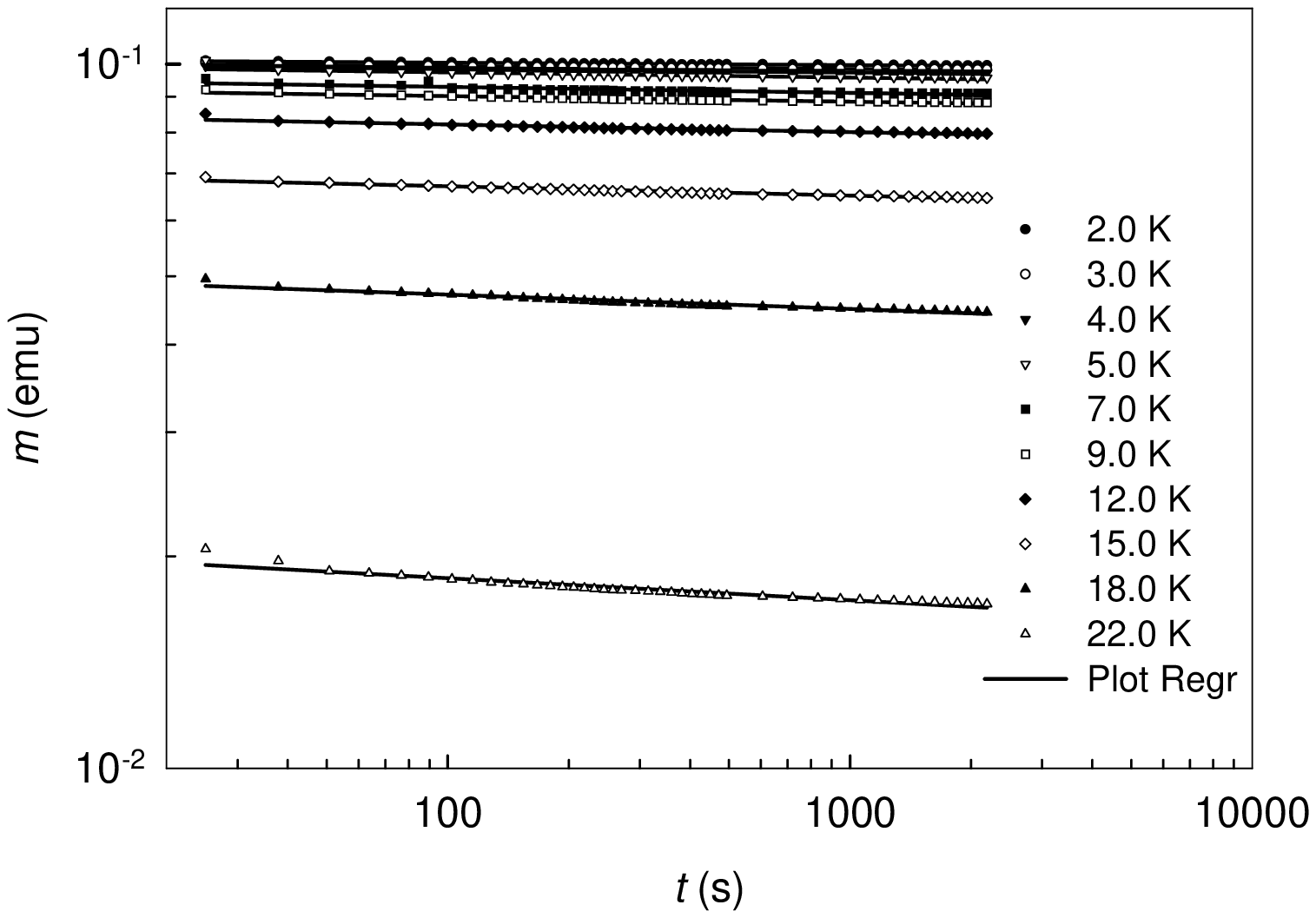}\caption{\label{f1} Magnetic relaxation data at magnetic field $\mu_0H=2.0$ T for 10 different temperatures.}
\end{figure}
Figure~\ref{f1} presents magnetic relaxation data of the sample in the $\log$-$\log$ scale at 2.0 T for 10 fixed temperatures from 2.0 to 22 K.  All curves in Fig.~\ref{f1} approximately linear in the $\log$-$\log$ scale, except for the data in the initial time stage as illustrated in Fig.~\ref{f1}, which could mainly correspond to two reasons: the linear simulation (in the $\log$-$\log$ window) being too simple for simulating each $m(t)$, and reconfiguration of the flux line gradient, caused by a suddenly applied magnetic field. For a simple approach  as shown with linear fits in Fig 1, we however can omit the data in an initial stage, and simulate the data using Zeldov model, $\ln m \stackrel{\mu\to 0}{=}\ln m_0-\ln [1+(\mu T/U_0)\ln (t/t_0)]/\mu\stackrel{\mu\to 0}{\to}\ln m_0-(T/U_0)\ln(t/t_0)$, with $m_0=m(t_0)$ and $t_0=1$ s. The facility of this simulation is that leads to a simple relation $R=-d\ln m/d\ln t\approx T/U_{0}$ without  changing the qualitative conclusion of the energy barrier from a more precise simulation. 

\begin{figure} 
\includegraphics
[width=.6\columnwidth]
{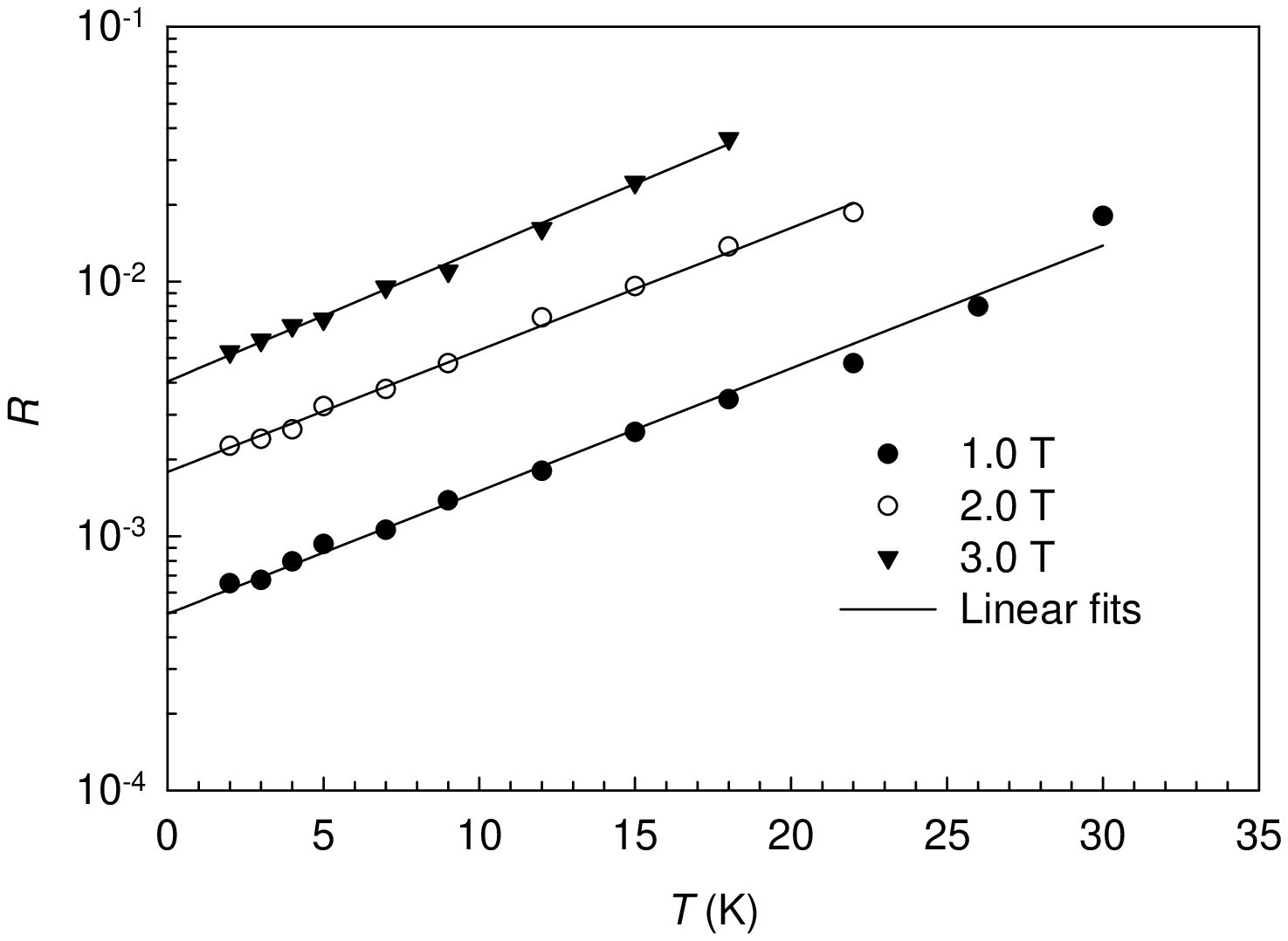}\caption{\label{f2} Temperature dependent relaxation rates at magnetic fields 1.0, 2.0 T and 3.0 T.}
\end{figure}
Figure~\ref{f2} presents the $R(T,H)$ curves of the MgB$_2$ sample at different magnetic fields for $t\ge 600$ s (for largely reducing the influence of the reconfiguration of the flux line gradient in the initial stage).  These data are strongly magnetic field dependent and show approximately linear relations in the linear-$\log$ window, resulting in that each $R$ curve is approximately simulated by a monotonic relation $R= R_0\exp(aT)$ as shown with solid lines, where $R_0$ is  the quantum creep at $T=0$ K, and $a$ the magnetic field dependent constant.  Note that this monotonic relation gives a remarkable comparison to that of a HTSC for which a broad peak or a plateau of a $R$ curve in the intermediate temperature range ($0.2 T_c < T < 0.8 T_c$) is observed \cite{Yesturun, Campbell, Civale}. $R(T,H)$s of a HTSC show monotonic reductions with decreasing temperature at low temperatures, where quantum tunneling of the vortices may play an important role for the magnetic moment decay. Considering the similar $R$ decreasing tendency over whole superconducting temperature regime for MgB$_2$, we may conclude that quantum tunneling exists through the regime.

According to the quantum tunneling theory \cite {Blatter,Blatter1,Blatter2} for the single vortex tunneling in the limit of strong dissipation, the decay rate $R(T)\approx \hbar/S(T)\approx e^2\rho_N(T)/\hbar L_c(T) =Q(T)[j_c(T)/j_0(T)]^{1/2}$, where $S$ is the saddle-point solution (bounce) of the Euclidean action, $L_c= \xi\sqrt{j_0/j_c}$ the collective pinning length along the magnetic field direction, $Q=e^2\rho_N/\hbar \xi$ the dimensionless quantum sheet resistance, $\rho_N$ the normal state resistivity, $\xi$ the coherence length, $j_0$ the depairing current density, and $j_c$ the critical current density. As a consequence, we can assume that the relaxation rate is the sum of two contributions corresponding to the purely thermal contribution, and the additional quantum tunneling contribution respectively. Hence, the relaxation rate is: 
\begin{equation}R=-\frac{d \ln m_q}{d \ln t} \approx \frac{T}{U_0+\mu T\ln(t/t_0)} + R_0 f(T) \label{3}, \end{equation} 
where $m_q$ is the magnetic moment using the quantum correction, $R_0=R(T=0)$ and $f(T)$ accounts for the temperature dependence of the quantum contribution, relating to the temperature dependent $S(T)$. Integrating Eq.(\ref{3}), we obtain the magnetic moment for the quantum correction:
\begin{equation} m_q(t)\approx \frac{m_0}{t^{R_0f}[1+(\mu T/U_0)\ln(t/t_0)]^{\frac{1}{\mu}}}\label {4}. \end{equation}
Considering a typical experimental time $t < 10^5$ s and a typical decay rate $R_0\le 0.01$ with $0\le f(T)\le 1$, one will find the additional factor $t^{R_0f}\le 10^{0.05}\approx 1.1$ leading to $m(t)\approx m_q(t)$ and $d \ln m/d \ln t\approx d \ln m_q/d \ln t$ in a broad temperature regime and a limited time scale. 

\begin{figure} 
\includegraphics
[width=.6\columnwidth]
{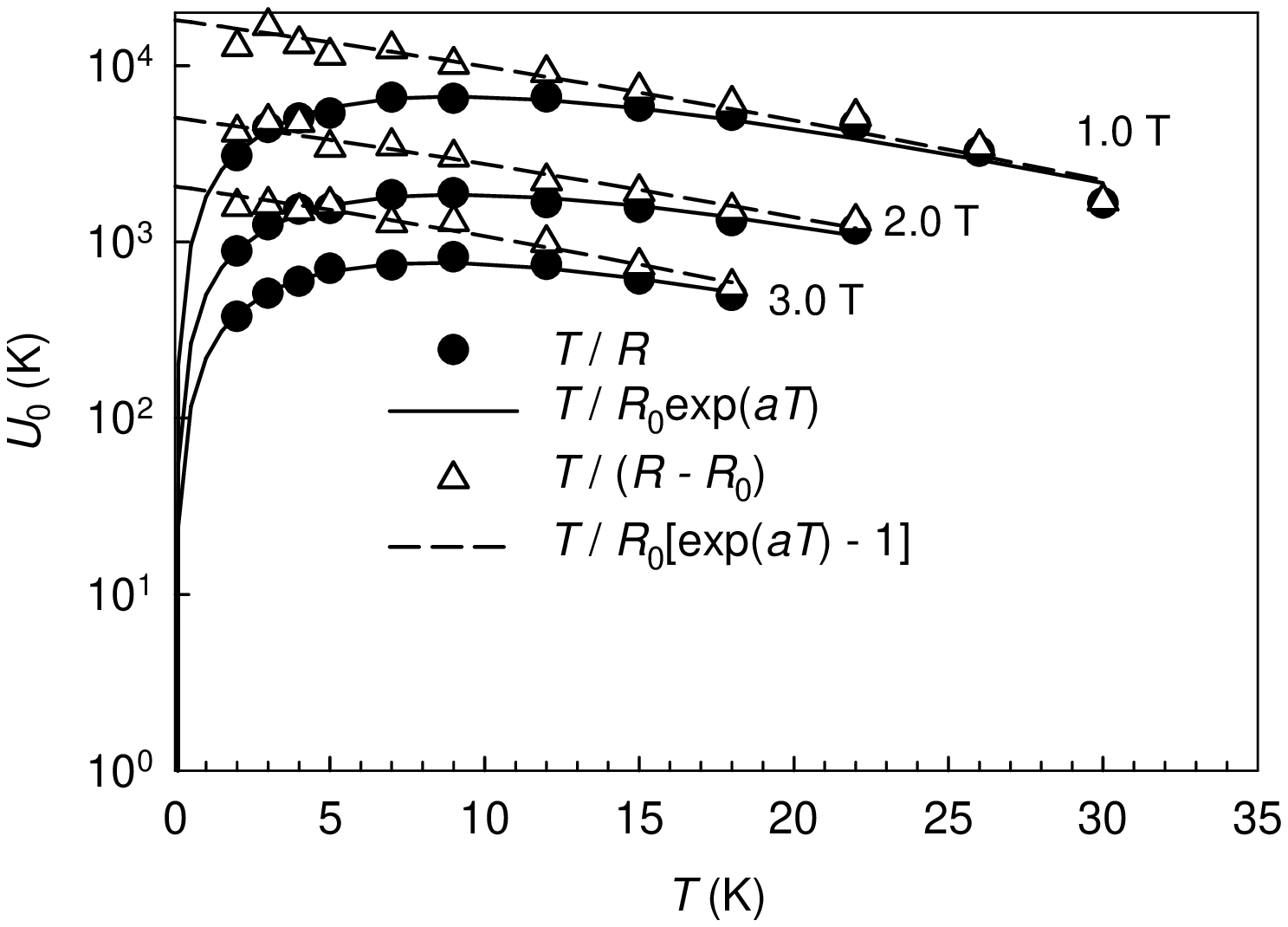}\caption{\label{f3} Energy barriers determined by different processes. The solid circles, and open triangles correspond to the $U_0$ determined by $U_0=T/R$, and $U_0=T/(R-R_0)$ respectively. The solid lines, and dash lines represent the relations $U_0=T/R_0\exp(aT)$, and $U_0=T/R_0[\exp(aT)-1]$,  respectively.}
\end{figure}
The interest is the comparison of the barrier evaluations between Eq.(\ref{2}) and Eq.(\ref{3}). The solid circles in Figure \ref{f3} show the $U_0(T,H)$ data determined using Eq.(\ref{2}) with $\mu=0$ ($U_0=T/R$). The solid lines represent the corresponding fits of the data using the relation $U_0=T/R=T/R_0\exp(aT)$. From the fits, it is easy to find that each $U_0(T,H)$ curve shows a maximum value (peak) in the intermediate temperature regime and $U_0(T\to)\to 0$ in the very low temperature regime. The tendency of $U_0(T\to 0)\to 0$ is unreasonable and contrasts sharply with the other superconducting properties, $e.g.$ the lower the temperature, the larger the irreversibility field and the larger the critical current. For the barrier evaluations with Eq.(\ref{3}), the temperature independent contribution, corresponding to $f(T)\equiv 1$,  is taken \cite {Blatter1,Blatter2}, leading to the barrier $U_0=T/(R-R_0f) =T/(R-R_0)$ as shown in Fig.~\ref{f3} with the open triangles. Remarkably, using the approach $R=R_0\exp(aT)$,  we find that each $U_0=T/R_0[\exp(aT)-1]$ curve (dash line) in Fig.~\ref{f3} monotonically increases with decreasing temperature and reaches its maximum value $U_0=1/R_0a$ at $T=0$ K, which notably contrasts to the corresponding $U_0=T/R=T/R_0\exp(aT)$ curve ($U_0(T\to 0)\to 0$).  For the extrapolations, one will find that the smaller the $R_0$, the larger the quantum correction in low temperature; besides, one will find that quantum correction is relatively small in high temperature where thermal effects are dominant. As a result, we conclude that a quantum correction of the relaxation rate is necessary for properly evaluating the energy barrier.

According to the quantum mechanics, the tunneling probability of vortices depends on  the barrier height. Since the height is temperature dependent, the $f(T)\equiv 1$ may not be in reality.  Therefore, we assume that both $R(T)$ and $f(T)$ can be simulated by polynomials, 
\begin{eqnarray}
R=R_0+\sum_{n\ge 1} R_nT^n\label{5}\end{eqnarray}
and 
\begin{eqnarray}
f=1+\sum_{n\ge 1}f_nT^n\label{6}, \end{eqnarray} 
where  $R_n$ and $f_n$ are polynomial coefficients, and $n$ is the integer for both polynomials.  
Taking Eqs.(\ref{5}) and (\ref{6}) to Eq.(\ref{3}), we have
\begin{eqnarray}
U_0(T) &&\approx {T}(R-R_0f)^{-1}\cr &&\approx (\sum_{n\ge 1} R_nT^{n-1}-R_0\sum_{n\ge 1}f_nT^{n-1})^{-1},\label{7}
\end{eqnarray}
leading to
\begin{eqnarray}
U_0(T= 0)\approx (R_1-R_0f_1)^{-1}.\label{8}
\end{eqnarray}
In general, we have
\begin{eqnarray}(R-R_0)/T=\sum_{n\ge 1} R_nT^{n-1}\ge 0\label{9}.
\end{eqnarray} 
Using the boundary conditions: $0\le f(T)\le 1$, $f(T=T_c)= 0$ and $f(T=0)=1$ , we determine
\begin{eqnarray}
(f-1)/T=\sum_{n\ge 1}  f_nT^{n-1}\le 0\label{10}. 
\end{eqnarray} 
Using Eqs.(\ref{9}) and (\ref{10}) to Eq. (\ref{7}), one will find that  $U_0(T)$ increases with decreasing temperature and reaches its maximum at $T=0$.

In summary, we have measured the temperature dependence of the magnetic relaxation rates of the MgB$_2$ superconductor in three magnetic fields. Based on the theory of quantum tunneling of vortices and the magnetic relaxation data in this work, a quantum correction of the relaxation rate, followed by the correction of the magnetic moment, is proposed for evaluating the energy barriers of the vortex motion.  Using this correction, we find that $U_0$, which sharply contrasts to the result without using the correction, increases with decreasing temperature and approaches a maximum at $T=0$ K. 

This work has been financially supported by the National Science Foundation 
of China,  and PAI 4/10 (Belgium).

 \end{document}